# An improved semiclassical method with reference potentials

**N.N.Trunov**

trunov@ vniim.ru

*D.I.Mendeleyev Institute for Metrology*

*Russia, St.Peterburg. 190005 Moskovsky pr. 19* introduce

(Dated: November 15, 2013)

**Abstract.** We approximate given potentials by means of the specially introduced reference potentials. On the one hand their parameters may be easily found from the usual WKB integral for the given potential; on the other hand they allow a simple determination of the first correction to the usual quantization condition.

Such reference potentials are introduced in special variables as some Pade approximants which lead usually to a very good results for various problems.

Thus we raise the accuracy of the WKB method for a broad set of potentials without cumbersome calculations. Higher order corrections as well as the exact condition is also considered.

### 1. Introduction

The semiclassical approximation with all its variations is still one of the main, simplest and most universal methods for solving problems where the exact solutions are cumbersome or unknown – not only in the quantum theory but also in various fields of science. Advantages of these methods as compared with the computers numerical ones are not only their simplicity but also a clear physical meaning of each stage of calculation. In particular, the explicit dependence on a given potential and other parameters can be easily traced. In its turn, it allows to introduce new physical conceptions and parameters. For example, an effective quantum number for centrally symmetric problems was constructed and used for many purposes[1].

The usual semiclassical quantization condition is only the first term of some series expansion with cumbersome terms including many differentiations of a given potential. That is why higher terms are almost always omitted.

The great advantage of the semiclassical method is its simplicity. We have already developed several ways to make more precise this method without any serious additional calculations [2, 4]. Hereafter we propose to approximate given potentials by means of the specially introduced reference potentials. On the one hand their parameters may be easily found from the usual WKB integral for the given potential; on the other hand they allow a simple determination of the first correction to the usual quantization condition

We represent these reference potentials by means of the Pade approximants in special variables introduced earlier. It is well known that such approximants usually lead to an unexpectedly good accuracy for many quite different problems. Simple algebraic structure allows to find necessary parameters and additions to the quantization condition without any cumbersome calculations.

## 2. General equations

Without loss of generality we can write the equation determining energy levels $\varepsilon_n$ in a potential well $V(x)$ in the following form:

$$\Phi(\varepsilon) = \frac{1}{\pi\beta}\int\sqrt{\varepsilon - V}\,dx = n + \tfrac{1}{2} + \delta \qquad (1)$$

with the unknown yet function $\delta$ which ensures the exact spectrum. Here $n = 0,1,2...$, the integral is taken between turning points and

$$\beta^2 = \hbar^2/2m \qquad (2)$$

The common popular semiclassical condition corresponds to $\delta \equiv 0$. It is known that such condition is exact only for few potentials. A power series expansion

$$\delta = \sum_{k=1}^{\infty} \delta_k$$
$$\delta_k = f_k(\varepsilon) \beta^{2k-1} \qquad (3)$$

is known [5] but is not practically used since $\delta_k$ are very cumbersome for $k > 1$,

$$\delta_1 = \frac{\beta}{24\pi} \frac{\partial^2}{\partial \varepsilon^2} \int \frac{dx}{\sqrt{\varepsilon - V}} \left(\frac{dV}{dx}\right)^2. \qquad (4)$$

For all potentials which may be expressed by means of an auxiliary function $s(x)$ as [3, 5]:

$$V(x) = A^2 s^2 + Bs + C$$
$$\sigma \equiv \frac{ds}{dx} = a_2 s^2 + a_1 s + a_0 \qquad (5)$$

the quantization condition (1) is exact if we put [5]

$$\delta = \frac{2\delta_1}{1 + \sqrt{1 + 16\delta_1^2}}; \qquad (6)$$

.

For all the potentials (5) and only for them

$$\gamma(\varepsilon) = \frac{d\delta_1}{d\varepsilon} \equiv 0, \qquad (7)$$

so that $\delta$ (6) is also independent on $\varepsilon$. The value of $\delta_1$ (4),

$$\delta_1 = \frac{\beta a_2}{8A} \qquad (8)$$

is invariant under the transformation $V \to V - C$, $s \to s + const$, in many cases it is convenient to choose $V = A^2 s^2$ (with the same $a_2, A$).

We can also obtain from (6) all higher corrections as a series expansion in powers of $\delta_1$:

$$\delta_3 = -4\delta_1^3 \tag{9}$$

and so on for all the potentials (5). Though such expansion is impossible if $16\delta_1^2 \geq 1$, our $\delta$ (6) remains exact at any $\delta_1$. The maximum absolute value $|\delta| = 1/2$ is reached when $|\delta_1| \to \infty$.

### 1. Reference potentials

As we have seen above, the distinction between the WKB ((1) with δ = 0) and exact spectra is fully concentrated in the phase addition δ (3). In many cases it is sufficient to find only $\delta_1$ - when $\delta_1$ is small enough or δ may be approximately calculated from (6) for potentials near to (5). Thus it shall be useful to have a simplified way to find $\delta_1$ without operations of (4) (not suitable at all for potentials given numerically).

Note that (5) is not a given *ad hoc* class but it embraces all the potentials used in handbooks as model and reference ones for the semiclassical approximation as well as for exact solutions [3 5]. (Excepting those potentials for which spectra are roots of some transcendental functions). This class embraces the potentials related to the factorization method as well as to the supersymmetric theory with an additive parameter.

Thus we expect that all or almost all physically actual potentials have in their vicinity (in some suitable sense) potentials satisfying to (5). The simplest criterion of this vicinity may be connected with smallness of γ (7).

In order to calculate $\delta_1$ we approximate given potentials by means of the specially introduced *reference* potentials. On the one hand they allow a simple determination of the first correction to the usual quantization condition; on the other hand, their new parameters may be found from the usual WKB integral $\Phi$ (1). Thus we raise the accuracy of the WKB method for a broad set of potentials without cumbersome calculations.

It is worth to mention that in general case each $\delta_n$ as well as $\delta$ itself may be expressed as some functional of $\delta_1(\varepsilon)$ and $\Phi(\varepsilon)$. It follows from the fact proved earlier [6]: each potential is fully determined by $\delta_1(\varepsilon)$ and $\Phi(\varepsilon)$. Thus the same pair $[\delta_1(\varepsilon), \Phi(\varepsilon)]$ determines the spectrum and in particular $\delta(\varepsilon)$.

Dependence of $\delta_n(\varepsilon)$ on $\delta_1(\varepsilon)$ and $\Phi(\varepsilon)$ may be certainly very cumbersome. But one may expect that simple approximate expressions exist for the potentials which are similar to (5).

**3. The Pade approximants for a given potential and determination of their parameters**

The reference potentials we introduce by means of some generalization of $\sigma$ (5), namely as some [2,2]- Pade approximant, i.e. the ratio of two quadratic polynomials. It is known that such Pade approximation leads usually to an unexpectedly good accuracy of results.

In the present paper we treat potentials $V$ with a sole parabolic minimum
$$V(x) \to k^2 x^2, \quad x \to 0, \qquad (10)$$
and introduce a new function $s(x)$ such that $A = 1$ in (5):
$$V(x) = s(x)^2. \qquad (11)$$
For a potential well with the height $U$, so that
$$V(x) \to U \quad at \quad |x| \to \infty. \qquad (12)$$
it means that
$$dV/dx = ds/dx = 0 \quad at \quad V = s^2 = U. \qquad (13)$$

Thus our suitable

$$\sigma = ds/dx = k(1- cs^2)/(1 + bs + gs^2), \qquad c = 1/U. \qquad (14)$$

with unknown yet $b$ and $g$. Of course no zeros within the interval [-1,1] are affordable.

In order to find these parameters we represent $\Phi(\varepsilon)$ by means of $\sigma$ (14):

$$\Phi(\varepsilon) = \frac{1}{\pi\beta}\int \frac{\sqrt{\varepsilon - s^2}}{\sigma} ds \qquad (15)$$

and divide it into two parts: $\Phi^+$ for $x > 0$ and $\Phi^-$ for $x < 0$ so that

$$\Phi(\varepsilon) = \Phi(\varepsilon)^+ + \Phi(\varepsilon)^-. \qquad (16)$$

Since $1/\sigma$ is linear in $bs$ (odd) and $gs^2$ (even), $b$ and $g$ may be determined simply and independently from the following expressions:

$$\Phi^+ + \Phi^- = \frac{2}{\pi\beta k}\int_0^{\sqrt{\varepsilon}} \frac{\sqrt{\varepsilon - s^2}}{1 - cs^2}(1 + gs^2)ds, \qquad (17)$$

$$\Phi^+ - \Phi^- = \frac{2b}{\pi\beta k}\int_0^{\sqrt{\varepsilon}} \frac{\sqrt{\varepsilon - s^2}}{1 - cs^2} s\, ds, \qquad (18)$$

where $\Phi^+$ and $\Phi^-$ are calculated for the given (maybe numerically) potential.

For $\varepsilon = U$, when a new level appears, we get from (17), (18)

$$g = 2\beta k\, (\Phi(U)^+ + \Phi(U)^-)/U^2 - 2/U, \qquad (19)$$
$$b = \pi\beta k(\Phi(U)^+ - \Phi(U)^-)/(2U^{3/2}). \qquad (20)$$

### 4. Calculation of the first addition $\delta_1$

Being expressed by means of $\sigma$, the first addition is

$$\delta_1 = \frac{\beta k}{6\pi} \frac{d^2}{d\varepsilon^2} \int_{-\sqrt{\varepsilon}}^{\sqrt{\varepsilon}} \frac{s^4(1-cs^2)ds}{\sqrt{\varepsilon-s^2}(1+bs+gs^2)} \qquad (21)$$

For symmetric potentials $b = 0$, then from (21)

$$\delta_1 = -\frac{\beta k(c+g)}{8(1-\varepsilon g)^{5/2}} \quad . \qquad (22)$$

In the opposite case, when $g = 0$,

$$\delta_1 = -\frac{\beta k(c-b^2)}{8(1+\varepsilon b^2)^{5/2}} \qquad (23)$$

and its absolute value decreases with increasing $\varepsilon$. The result of (21) may be represented as a series expansion:

$$\delta_1(\varepsilon) = -\beta k(c+g-b^2)/8 + 5\beta k\varepsilon(cg+g^2+b^4-3b^2g-cb^2)/16 + \ldots \qquad (24)$$

For the sake of completeness we write down the basic form of the potential well

$$V = U \tanh^2 px, \qquad (25)$$

for which $b = g = 0$ in (14), all the equations of Sect. 2 are exact,

$$k = p\sqrt{U}, \qquad (26)$$

and the first addition (as well as higher ones) is independent on $\varepsilon$:

$$\delta_1 = -\frac{\beta p}{8\sqrt{U}}, \qquad \delta_3 = -4\delta_1^3 \qquad (27)$$

and so on.

This dependence of $\delta_1$ on $U$ is a general property though $b$ and $g$ may be smooth functions of $\varepsilon$ in our cases. Note that we do not need the explicit form $V(x)$ with $\sigma$ (14).

## 5. The total phase addition $\delta$

For basic potentials, e.g. (25), the value of $\delta$ is exactly determined by means of $\delta_1$, see (6). This connection must be approximately valid for all our potentials similar to the basic ones. But there is an interesting case when (6) is exact for any form of the potential well. It is known that each one-dimensional potential well satisfying the above condition (16) has a bound state even its depth tends to zero. As it follows from (27),

$$\delta_1 \to -\infty \qquad at \ U \to 0. \qquad (28)$$

Accordingly to (6),

$$\delta \to -1/2 \qquad (29)$$

and compensates ½ in the right side of (1), so it becomes equal to zero for the lowest level $n = 0$. But if $U \to 0$ then obviously the level energy $\varepsilon \leq U \to 0$ so that $\Phi \to 0$ in (1). Thus namely $\delta$ (6) ensures the validity of the quantization condition even for the lowest level. Another applications of (6) for new interpolations of the total $\delta$ may be found in [ 2 ].

As it is obvious from (1), (25), the total number of the eigenstates $N$ is proportional to $\sqrt{U}$, so that $\delta_1 \ll 1$ if $N$ is great enough, and $\delta \approx \delta_1$. This property remains valid for all similar potentials with small $b \neq 0$ and/or $g \neq 0$.

## 6. Discussion

The proposed approach allows many variations and improvements. E.g. we can take into account an important property of the state density

$$P = dN(\varepsilon)/d\varepsilon,$$

where $N(\varepsilon)$ is the total number of the eigenstates with energies not exceeding $\varepsilon$. As it is seen from (1), in main approximation $P$ is proportional to $d\Phi/d\varepsilon$. For a finite potential well (12) with $V$ belonging to the class (5) or close to it,

$$P(\varepsilon) \approx const\ (d\Phi/d\varepsilon) \to \infty \quad if \quad \varepsilon \to U. \tag{30}$$

Suppose $\sigma$ for our potential is (14) with $g = 0$ and unknown yet $c$ and $b$. Differentiating both sides of (15) on $\varepsilon$ and calculating the simple integral we obtain

$$c = \left\{ \left[ \frac{1}{2k\beta(d\Phi/d\varepsilon)} \right]^2 - 1 \right\} \frac{1}{\varepsilon}. \tag{31}$$

At the same top of our potential well, using (30), we get in accordance with the previous results

$$c = -1/U, \quad \varepsilon = U,$$

so that $\sigma = 0$ and in fact our potential does not grow more.

Of course, we can use the same expression (18) for oscillator-like potentials with no limiting condition like (12),

$$V(x) \to \infty \quad at \quad |x| \to \infty \tag{32}$$

and a finite $P$ at all $\varepsilon$.

We have already mentioned that studied above potentials must be similar to the basic ones, for which $b = g = 0$ and $\delta_1$ (as well as $\delta$) does not depend on $\varepsilon$. Respectively this similarity means that: 1) $b^2$ and $|g|$ are smaller (or better considerable smaller) than $|c|$ and 2) all the coefficients, which are constant for the basic potentials, change "adiabatically", e.g. in

$$d(c\varepsilon)/d\varepsilon = c + \varepsilon dc/d\varepsilon$$

the second term must be considerable smaller than the first one:

$$|dc/d\varepsilon| << |c/\varepsilon|. \tag{33}$$

Applications of the above method will be presented in anoter paper.